\newif\ifarxiv
\arxivtrue

\ifarxiv
\documentclass[%
 amsmath,amssymb,
 aps,
]{revtex4-2}

\else

\fi

\usepackage[T1]{fontenc}
\usepackage[postscript]{ucs}

\ifarxiv
\else 
\fi

\usepackage{graphicx}
\usepackage{epstopdf, epsfig}
\usepackage{verbatim}
\usepackage{xfrac}
\usepackage{enumitem}
\usepackage{mathtools}
\usepackage{bm}
\usepackage{color}
\usepackage[hidelinks]{hyperref}
\usepackage{cleveref}
\usepackage{mdframed}
\usepackage{booktabs}
\usepackage{wrapfig}
\usepackage[table,x11names]{xcolor}
\usepackage{nicefrac}
\usepackage{subcaption}
\usepackage{cases}
\usepackage{soul}
	\setstcolor{red}
\usepackage{siunitx}

\newcommand{\yieldstress}{\tau_y}
\newcommand{\Bin}{\text{Bi}}

\renewcommand{\vector}[1]{\mathbf{#1}}
\newcommand{\vu}{\vector{u}}
\newcommand{\vv}{\vector{v}}
\newcommand{\vx}{\vector{x}}
\newcommand{\vzero}{\vector{0}}
\newcommand{\vn}{ \hat{\vector{n}} }

\newcommand{\ttau}{\bm{\tau}}
\newcommand{\tdel}{\bm{\delta}}
\newcommand{\tsigma}{\bm{\sigma}}
\newcommand{\tI}{\mathbb{I}}

\newcommand{\parD}[2]{\frac{\partial #1}{\partial #2}}
\newcommand{\divergence}[1]{\text{div}\,#1}

\usepackage{color}
\usepackage{MnSymbol}
\usepackage{layouts}
\definecolor{orange}{rgb}{1.0,0.4,0.0}

\ifarxiv

\begin{document}

\ifarxiv
\preprint{APS}
\title{Squirmer locomotion in a yield stress fluid}

\author{Patrick S. Eastham}
\affiliation{Department of Psychology, Florida State University,
Tallahassee, FL 32311, USA}
\author{Hadi Mohammadigoushki}
\affiliation{Department of Chemical and Biomedical Engineering, FAMU-FSU College of Engineering, Tallahassee, FL, 32310, USA}
\author{Kourosh Shoele}
 \email{kshoele@eng.famu.fsu.edu}
\affiliation{Department of Mechanical Engineering, FAMU-FSU College of Engineering, Tallahassee, FL, 32310, USA}

\else 
\fi

\begin{abstract}
An axisymmetric squirmer in a Bingham viscoplastic fluid is studied numerically to determine the effect of a yield stress environment on locomotion. The nonlinearity of the governing equations necessitates numerical methods, which is accomplished by solving a variable-viscosity Stokes equation with a Finite Element approach. The effects of stroke modes, both pure and combined, are investigated and it is found that for the treadmill or ``neutral'' mode, the swimmer in a yield stress fluid has a lower swimming velocity and uses more power.
{However, the efficiency of swimming reaches its maximum at a finite yield limit. In addition, for higher yield limits, higher stroke modes can increase the swimming velocity and hydrodynamic efficiency of the treadmill swimmer. The higher-order odd-numbered squirming modes, particularly the third stroke mode, can generate propulsion by themselves that increases in strength as the viscoplastic nonlinearity increases {till a specific limit}. These results are closely correlated with the confinement effects induced by the viscoplastic rigid surface surrounding the swimming body, showing that swimmers in viscoplastic environments, both biological and artificial, could {potentially} employ {other} non-standard swimming strategies to optimize their locomotion.}
\end{abstract}

\ifarxiv
\maketitle
\fi

\section{Introduction}
\label{sec:intro}

Swimming microorganisms live in a variety of fluid environments. While locomotion in Newtonian environments at low Reynolds number is relatively well-established \citep{lauga2016bacterial}, swimming in complex fluids is an area of active research due to the relevance of biological fluids, which often have highly non-Newtonian properties. Understanding the role of the non-Newtonian fluid properties on swimming performance is important for the design of artificial swimmers \citep{bunea2020recent, wu2020medical, tsang2020roads}.

\emph{Helicobacter pylori} is an infectious microorganism that moves through dense, gel-like gastric mucus by actively changing its local environment to enable locomotion \citep{bansil2013influence}. A gastric mucus and other gel-like materials, unlike Newtonian fluids, can have solvent-gel transition and, in general, do not follow a single canonical fluid model. Because the fluid region near \emph{H.~pylori} is understood to be easier to swim through than the outer gel, one model of this situation combines an inner Stokes fluid with either an outer Stokes fluid with a different viscosity \citep{reigh2017two} or an outer Brinkman fluid \citep{nganguia2020squirming}. 
{Similar to many other researches, these studies used an idealized model of a swimmer as a squirmer  \citep{lighthill1952squirming,blake1971spherical,pedley2016spherical}. 
The squirmer model has been employed to study the locomotion in various complex fluids \citep{lauga2009life,zhu2012self,li2014effect,datt2017active,pietrzyk2019flow,binagia2020swimming,van2022effect}.
When both the swimmer and the inter-fluid boundary are spherical or ellipsoids, these models are analytically tractable and consequently can be used to produce exact results for a wide range of  parameters.}   However, these models suffer from the deficiency that the confinement is a fixed, independent parameter. A more realistic model would allow the confinement radius to be induced by the swimming body itself, either through surface kinematics or chemical reaction. In this paper, we ignore any chemistry-induced changes to the fluid and focus purely on hydrodynamic effects. The role of chemistry-induced fluid changes on locomotion is discussed in \citet{mirbagheri2016helicobacter}.

To study the above effects, a specific model is needed that intrinsically links the fluidity in the near-field to the fluid boundary conditions, i.e. the degree of confinement is not an independent parameter. This will lead to more natural results as would be encountered by a typical swimmer such as \emph{H.~pylori}. One canonical model that achieves this goal is a Bingham viscoplastic fluid subject to yield stress. This means that the material behaves as a plastic solid -- capable of rigid translation but not deformation -- when the interior stress is below a certain value, while it is allowed to deform when the stress exceeds this value. Common examples include hair gel and toothpaste, where the material can maintain its shape indefinitely under low shear but flows easily once adequate shear is applied. While in their most complex state, viscoplastic models can include the effect of shear-thinning or -thickening, i.e.~the Herschel-Bulkley model; here, we use a Bingham fluid model \citep{saramito2016complex} for simplicity while retaining the key viscoplastic component relevant to our biological motivation. Previous studies on swimming in Bingham fluids have been confined to two dimensions \citep{hewitt2017taylor, supekar2020translating} or for slender bodies \citep{hewitt2018viscoplastic}. Here, we extend these studies and consider an idealized steady spherical squirmer that employs tangential motions on its boundary.

{A closely related problem is the locomotion in shear-thinning and viscoelastic fluids \citep{datt2015squirming,elfring2015theory, li2021microswimming, wu2022formation}. By employing the squirmer model, it is shown that while a swimmer slows down in a shear-thinning fluid, they could have higher swimming efficiency and their maximum swimming efficiency occurs at a particular boundary actuation rate \citep{nganguia2017swimming}. A similar conclusion is reached for a wide range of spheroidal shape swimmers and it is found that spheroidal swimmers compared to over spherical squirmers, can reach higher swimming velocity and energetic efficiency in
shear-thinning fluids \citep{van2022effect}. In addition, it is shown that the nonlinear rheology of the fluid can modify the efficient swimming gaits in non-Newtonian fluids compared to Newtonian fluids and allows non-motile surface actuation modes in Newtonian fluids to generate net motion in shear-thinning fluids \citep{pietrzyk2019flow}. 
The changes in the swimming performance are associated with the relative importance of local and non-local effects in shear-thinning fluids. Such effects are more complex in yield fluids as the flow properties undergo rapid transition across the yield surface, which itself is formed based on the swimmer geometry and mode of actuation. To date, it is still unclear how the hydrodynamics of the squirmer model and its swimming kinematics are modified in yield fluids. This constitutes the main objective of this work. 
}

The paper is organized as follows: the governing equations of a squirmer in a Bingham fluid are described in Section \ref{sec:model}. The computational model is described in Section \ref{sec:numerics}. Section \ref{sec:results} contains results for treadmill and higher-order modes. We conclude with Section \ref{sec:discussion} by describing what these results suggest about the nature of swimming in viscoplastic materials, particularly commenting on the possibility of different optimal swimming strategies that employ higher-order modes.

\section{Governing Equations}
\label{sec:model}

We employ the regularized Bingham viscoplastic Stokes flow model. The regularization is adequate to model the dynamics of a small squirmer, on the length scale of microns, swimming through a yield stress fluid. At these scales, diffusion results in a smooth transition between yielded and unyielded regions. Yet, for a sufficiently small smoothing parameter, as will be explained later, the yield surface approaches the sharp interface encountered in the singular Bingham formulation.

Mathematically, we consider an exterior flow problem of a steady squirmer swimming in an infinite viscoplastic fluid given by the regularized Bingham model \citep{saramito2016complex, saramito2017progress}, written in a nondimensional form as:
\begin{equation} \label{eq:viscoplastic}
    -\divergence{\ttau} + \nabla p = \vzero,\qquad \ttau = 2\eta(\dot{\gamma})\,\mathbb{D}(\vu),\qquad \divergence{\vu}=0
\end{equation}
where
\begin{equation} \label{eq:effviscosity}
    \eta(\dot{\gamma}) = 1 + \Bin(\dot{\gamma}^2+\epsilon^2)^{\nicefrac{-1}{2}}
\end{equation}
{is the effective viscosity which depends on the shear rate $\dot{\gamma}=|2\mathbb{D}|$.} Here, $p$ is the pressure, and  $\bm{\tau}$ and $\mathbb{D}$ are the deviatoric stress and rate-of-strain tensors, respectively, $\vu$ is the fluid velocity and $\epsilon$ is a small regularization parameter. {The tensor norm is defined as $|\tdel|^2\,=\,(\tdel\bm{:}\tdel)/2$ for any tensor $\tdel\,\in \mathbb{R}^{3\times\,3}$.} 

The Bingham number, $\Bin{}= \yieldstress R / (\mu_0 V)$, is employed as the key characteristic parameter of the non-dimensional equations, where $R$ is the length scale, taken to be the radius of the squirmer, $\yieldstress$ is the yield stress, $\mu_0$ is the characteristic (Newtonian) fluid viscosity, and $V$ is some characteristic velocity scale. Because we consider swimming problems where the swimmer might not achieve motion, we let $V=\sqrt{\mathcal{P}_N / \mu_0 R}$ where $\mathcal{P}_N$ is the power expenditure of the squirmer in the Newtonian case \citep{blake1971spherical,michelin2011optimal}. The yield surface is defined to be associated with  $|\ttau| = \Bin{}$. We assume perfect viscoplasticity and the rigid region is placed at $|\ttau| < \Bin{}$. 

\begin{figure}
    \centering
    \includegraphics[width=0.6\linewidth]{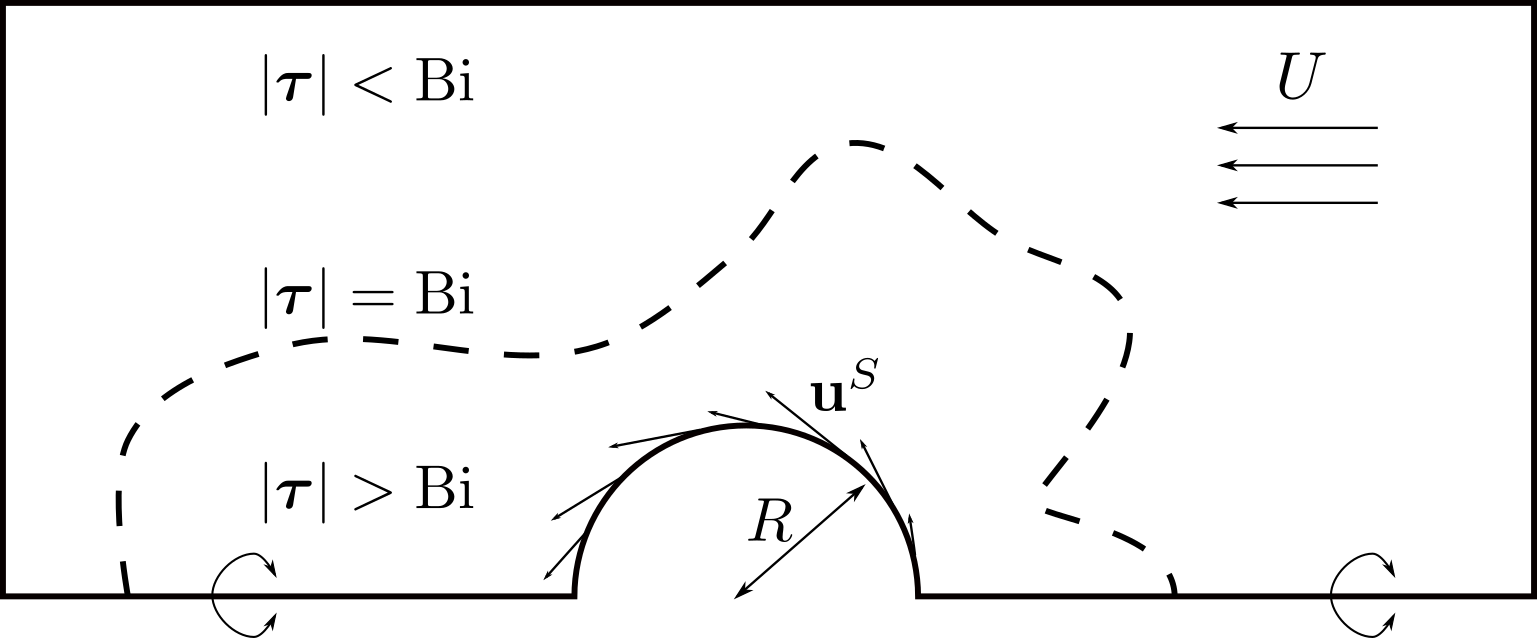}
    \caption{Schematic of axisymmetric squirmer in Bingham fluid environment. The diffuse yield surface (|$\ttau| = \Bin$) is given by the dashed line. The fluid region exterior to this line has negligible rate of deformation, i.e.~$\dot{\gamma}\ll 1$.}
    \label{fig:viscoplastic_schematic}
\end{figure}

The non-dimensional power $\mathcal{P}$ expended by the squirmer into the fluid is defined to be 
${\mathcal{P} = -\int_\mathcal{S}\vu^S\cdot(\vn\cdot\tsigma)\, dS}$
where $\tsigma = -p\tI + \ttau$ is the total stress on the body $\mathcal{S}$ and $\vn$ is the unit normal vector. The boundary conditions in the body attached coordinate are $\vu(|\vx|=1) = \vu_s$ and $\vu(\vx\to\infty) = -U \bm{e}_x$
where $\vu_s$ is the tangential squirming surface motion, defined in \eqref{eq:surfacestroke}, and $U$ is the translational swimming velocity of the squirmer in the $ \bm{e}_x$ direction.

{The surface motion $\vu_s$ consists of orthogonal ``stroke modes'' with the surface velocity of $\vu_s = u_\theta\vector{e}_\theta$, where $\vector{e}_\theta$ is the unit tangent vector on the surface of the sphere. The surface velocity magnitude $u_\theta$ is represented by \citep{blake1971spherical}
\begin{equation} \label{eq:surfacestroke}
u_\theta(\nu) = \sum\limits_{n=1}^\infty \beta_n K_n(\nu),\qquad K_n(\nu) = \sqrt{\frac{3}{n(n+1)}}\sqrt{1-\nu^2}P_n'(\nu)
\end{equation}
where $\nu = \cos\theta$. Here, $\theta$ is is the polar angle with respect to the swimming direction. For each orthogonal mode $n$, $P_n'(\nu)$ is the first derivative of the $n^{\text{th}}$-order Legendre polynomial. The formulation permits to fully characterize the swimming stroke using the coefficients $\beta_n$, and in the case of swimmer in the Newtonian fluid, to correlate the swimming velocity to the first squirming stroke mode as $U = \sqrt{2/3} \beta_1$. The choice of non-dimensionalization based on the rate of energy dissipation of Newtonian swimmer leads to a constraint relation between $\beta_n$ coefficients \citep{blake1971spherical,michelin2011optimal}. In particular, the stroke modes $\beta_n$ are subject to the following constraint,

\begin{equation}
\frac{2}{3}\beta_1^2\, + \,\sum_{n=2}^\infty\beta_n^2 \,=\, 1.
\end{equation}

While in the case of a classic Stokes fluid, this fixes the power expenditure of the squirmer \citep{blake1971spherical}, we will see that this is not the case for swimmers in a Bingham fluid. Nevertheless, this choice is made to enable us e to cross-compared different swimmers based on their energy expenditure and to study both swimming and non-swimming strokes.}

\section{Numerical Method}
\label{sec:numerics}

A continuous Galerkin Finite Element method (FEM) is used to numerically calculate the flow around a free-swimming steady squirmer \citep{eFEMpart2019}. The quadrilateral mesh is used, and the simulations with $\Bin{}=0$ are done with an outer radius of 20,000. All simulations with $\Bin{}>0$ are performed with an outer radius of 20; the presence of the yield surface means that velocity field deformations decay to negligible amounts rapidly, and so more elements can be included near the swimmer to capture the yield surface.
We use a Picard iteration scheme to solve the nonlinear system \eqref{eq:viscoplastic}.  {The procedure starts with $\Bin=0$ assumption and with the choice of larger regularization parameter. At each iteration, the fluid equation \eqref{eq:viscoplastic} can be interpreted as a variable-viscosity Stokes equation. The system is iteratively solved until convergence with an absolute difference tolerance of $10^{-6}$. Following the convergence, the yield surface is identified and mesh is resolved near the interface. Then, the solutions with higher $\Bin{}$ and smaller regularization parameter are recalculated and the procedure continues till the final convergence criteria of $10^{-6}$ is achieved.} The axisymmetric weak formulation at the $n^{\text{th}}$ iteration is
\begin{subequations} \label{eq:FEM}
\begin{align}
\int_D2\eta\left(\dot{\gamma}^{n-1}\right)\left[\mathbb{D}(\vu^n):\mathbb{D}(\vv) + \frac{u_r^n v_r}{r^2}\right]&r\,dS - \int_D{p^n(\nabla\cdot \vv) r}\,dS= 0, \\
\int_D q(\nabla\cdot \vu^n) r \,dS &= 0,
\end{align}
\end{subequations}
where $D$ is the axisymmetric domain, $r$ is the radial variable,  $(\vu^n, p^n)$ refers to unknowns to be solved for and $\dot{\gamma}^{n-1}$ is assumed to be known from the previous iteration's solution. The test functions $(\vv, q)$ are chosen to be Taylor-Hood Q$_2$--Q$_1$ shape functions. More details on the axisymmetric weak form for Stokes equation and the validation of the model can be found in \citet{eastham2020axisymmetric}. The yield surface is identified by applying a  bisection method radially to the scalar field $|\ttau| - \Bin{}$ for different $\theta$ values. 

\section{Results}
\label{sec:results}

We investigate a treadmill squirmer in a Bingham fluid and compare these results to an equivalent squirmer in a Newtonian fluid. Then we explore the combinations of surface modes and analyze the qualitative differences between even and odd modes.
Our goal is to understand the performance of the squirmer relative to its performance in a Newtonian fluid ($\Bin{}=0$). Therefore we define a normalized power as
$\mathcal{P}^* = \mathcal{P} / \mathcal{P}_N $
where the $N$ subscript indicates the power expended by the same squirmer in a Newtonian fluid environment. Because we consider ``purely mixing'' stroke modes with $\beta_1=0$, where $U_N=0$, we do not normalize the swimming speed $U$.

\subsection{Treadmill Squirmer}

\begin{figure*}
    \centering
    \begin{subfigure}[t]{0.32\textwidth}
        \centering
        \includegraphics[scale=0.85]{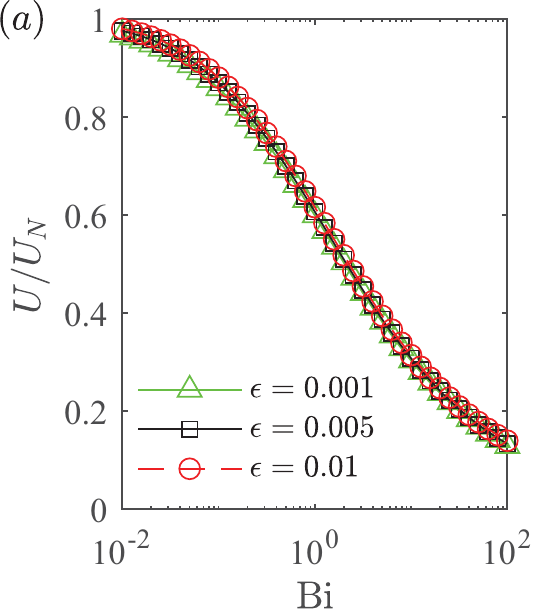}
    \end{subfigure}%
   \hspace*{\fill}
    \begin{subfigure}[t]{0.32\textwidth}
        \centering
        \includegraphics[scale=0.85]{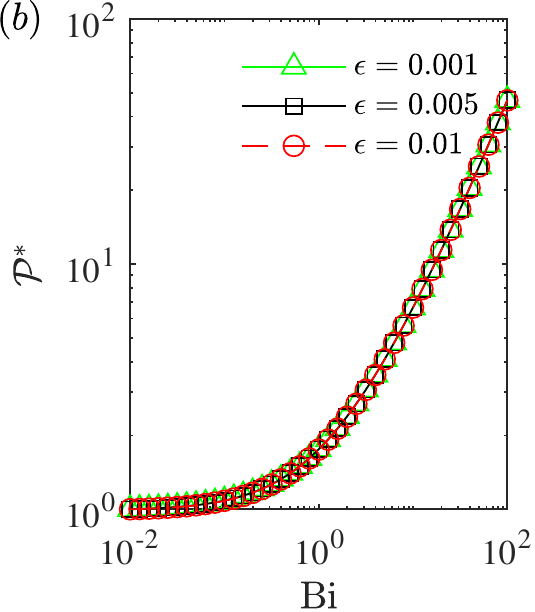}
    \end{subfigure}
   \hspace*{\fill}
    \begin{subfigure}[t]{0.32\textwidth}
        \centering
        \includegraphics[scale=0.85]{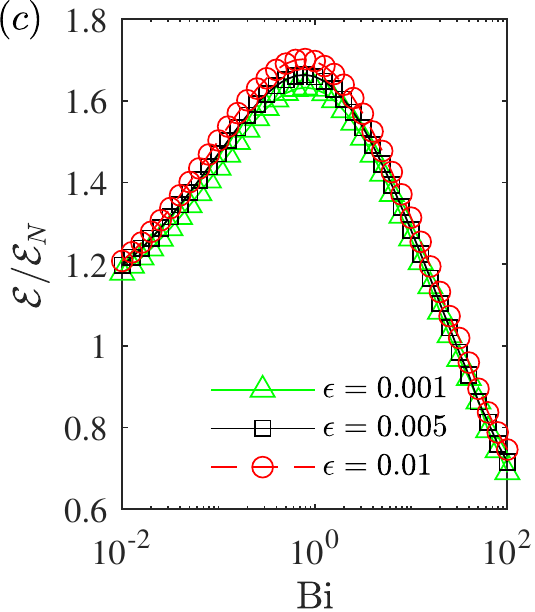}
    \end{subfigure}    
    \caption{(a) Swimming speed $U$,  (b) normalized power expenditure $\mathcal{P}^*$ and (c) ratio of the swimming efficiency compared with Newtonian efficiency $\eta_N$ of a squirmer with $\beta_1=\sqrt{3/2}$ for three  different regularization parameters of $\epsilon=0.001,0.005,0.01$. }
    \label{fig:squirmerPlots}
\end{figure*}

\begin{figure}
    \centering
    \begin{subfigure}[t]{1\textwidth}
        \centering
        \includegraphics[scale=0.75]{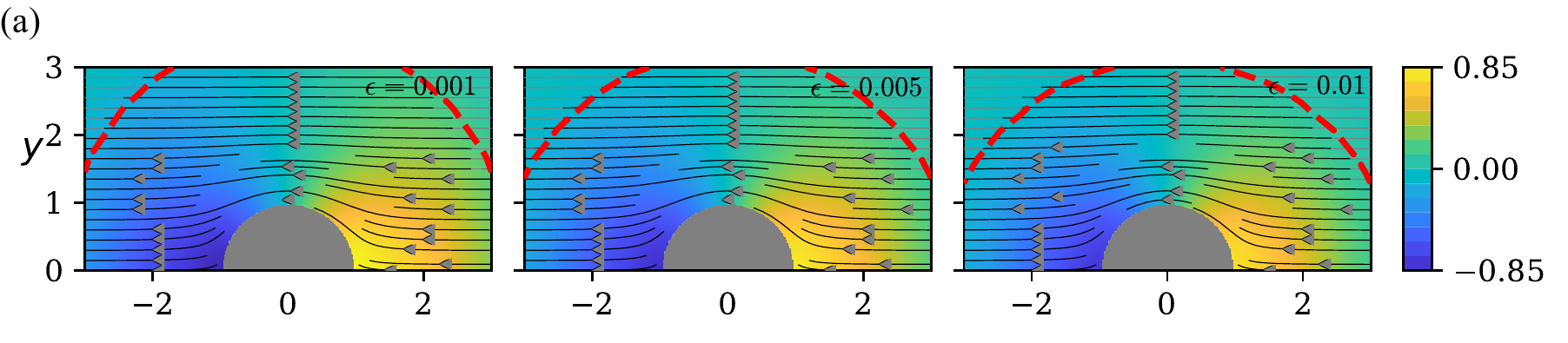}
    \end{subfigure}%
    
    \begin{subfigure}[t]{1\textwidth}
        \centering
        \includegraphics[scale=0.75]{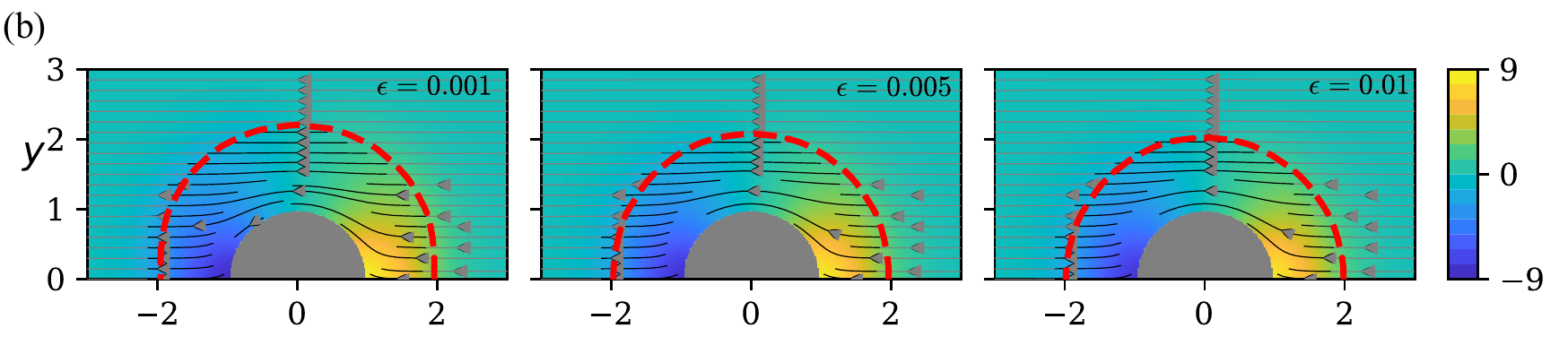}
    \end{subfigure}%
    
    \begin{subfigure}[t]{1\textwidth}
        \centering
        \includegraphics[scale=0.75]{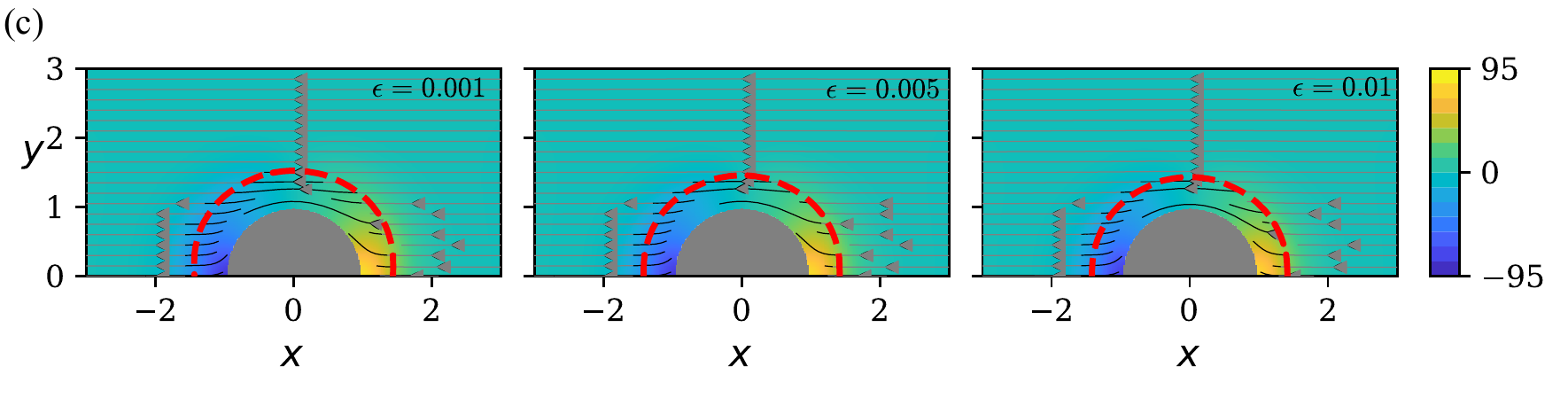}
    \end{subfigure}%
    \caption{ The velocity  (Streamlines) and pressure fields (colors) for $\Bin{}=0.1$ (a), $1.0$ (b) and $\Bin{}=10$ (c) and 3 regularization values of $\epsilon=0.001$, 0.005 and 0.01. The red-dashed line represents the yield surface.}
    \label{fig:treadmill_flow}
\end{figure}

We first simulate a treadmill squirmer, i.e.~$\beta_i = \sqrt{3/2}\delta_{i1}$ where $\delta_{ij}$ is the Kronecker delta. In Newtonian fluids, this is the only mode that generates locomotion while all other modes, generally termed ``mixing modes'', do not. We compute both the swimming speed and power expenditure for Bingham number $\Bin{}$. The value of  $\epsilon$ is selected to be sufficiently small such that it does not modify the trends observed here. In particular, it is found that the results are insensitive to the choice of $\epsilon$ if $\epsilon<0.01$ . This can also be seen from Figs \ref{fig:squirmerPlots} and \ref{fig:treadmill_flow}. In all cases, we found that the viscoplastic squirmer requires more power and swims slower than its Newtonian counterpart (Figs. \ref{fig:squirmerPlots}a,b). 
{The decrease of the swimming velocity is much faster at the intermediate $\Bin{}$ range of $\Bin{} \approx 1$. For higher $\Bin{}$ range,  the power expenditure increases exponentially with $\Bin{}$. To compare the efficiency of the swimmer in the viscoplastic and Newtonian fluids, we use the standard definition of swimming efficiency proposed by \cite{lighthill1975mathematical} and define the efficiency $\eta$ as,
\begin{equation}
\mathcal{E}\,=\,\frac{U\mathcal{D}}{\mathcal{P}}
\end{equation}
where the power needed to translate a rigid sphere at the same swimming speed, $U$, is compared to the power dissipation $\mathcal{P}$ in the viscoplastic fluid case. Here, $\mathcal{D}$ is the force required to drag a rigid sphere at the swimming speed in the same fluid medium and is calculated from separate simulations with the imposed swimming velocity. This definition of swimming efficiency has been widely adopted to characterize the efficiency of different swimmers \citep{stone1996propulsion,tam2007optimal,michelin2010efficiency,nganguia2018squirming} and is adopted here. A similar expression can be employed to find the efficiency of a swimmer in Newtonian fluid, $\mathcal{E}_N$. It is possible to show mathematically that for spherical squirmers with the pure treadmill motion, $\mathcal{E}_N=\frac{1}{2}$ \citep{lighthill1952squirming,blake1971spherical}. 

Figs. \ref{fig:squirmerPlots}c shows the efficiency of a squirmer as a function of $\Bin{}$. It is observed that for $\Bin{}<10$, the efficiency of the squirmer in the yield flow is higher than the Newtonian swimmer, attaining its maximum value of $\mathcal{E}/\mathcal{E}_N=1.4$ at $\Bin{}=1.0$. Only for very large yield limits, where the power increases exponentially, do we see lower efficiency than the Newtonian swimmer. The results reveal that the swimmer is most efficient at $\Bin{}=O(1)$.}

The flow fields are visualized in Figure \ref{fig:treadmill_flow}b where the yield surface is shown with the red line for three yield limits of $\Bin{}=0.1$(a), $\Bin{}=1$(b) and $\Bin{}=10$(c). {The streamlines outside the yield region are marked with a grey color.} Each column is for a particular regularization parameter $\epsilon$. First, it is found that the shape of the yield surface is independent of the choice of $\epsilon$ for sufficiently small values. As $\Bin{}$ increases, the yield surface shrinks closer to the surface and the plasticity becomes more dominant. The decreased swimming speed of treadmill squirmers with increased confinement is similar to results from other models that involve confinement \citep{reigh2017two,nganguia2020squirming}. The most drastic change is related to the pressure field. The pressure becomes larger with increasing $\Bin{}$, mainly because of the confinement effect from the yield surface. Mathematically, this is related to rapidly varying effective viscosity $\eta(\dot{\gamma})$  near the yield surface in equation \ref{eq:viscoplastic}. This effect, where pressure compensates for spatially varying viscosity, is seen in other complex fluid models with non-constant viscosity \citep{eastham2020axisymmetric,nganguia2017swimming}. 

{The results are different from the observed behavior in shear-thinning fluids \citep{nganguia2017swimming}. Contrary to shear-thinning fluids and how the swimming behavior changes with Carreau number (equivalently the ratio of strain rate and relaxation rate of the fluid medium), the velocity now continuously decreases with $\Bin{}$. Moreover, unlike shear-thinning fluids, where the efficiency is always larger than the Newtonian fluid, the swimmer efficiency in viscoplastic fluids can be higher or lower than the swimmer in Newtonian fluid due to rapid increase of power expenditure at large $\Bin{}$ cases. Nevertheless, both viscoplastic and shear-thinning fluids show local maxima in the swimming efficiency at intermediate values of $\Bin{}$ or Carreau number}. 

Although the swimming speed and power are clearly affected by viscoplasticity, the qualitative features of the velocity field are consistent between Newtonian and non-Newtonian cases. This will change when we include certain higher-order modes. As shown in the figures,  the results do not change with the choice of the regularization parameter $\epsilon$. Following the sensitivity analysis,  $\epsilon=0.005$ is fixed for the remaining of this paper.

\subsubsection{Surface Perturbation Analysis}
To form a conceptual bridge between the treadmill squirmers in Newtonian and viscoplastic fluids, we approximate the nonlinear viscoplastic environment as a confined linearly viscous environment around the swimmer. We follow \citet{reigh2017two} and derive an analytic solution to the problem of a spherical squirmer of radius $R$ swimming inside a Stokes fluids confined by a concentric viscoplastic yield surface, which is a perturbed spherical surface. In the following, we investigate how small surface perturbations of the outer surface affect the swimming velocity.

Mathematically, this problem is defined by considering the Stokes squirmer with boundary conditions in the reference frame:
\begin{equation}
    u_\theta\Big|_R = u_\theta^0\Big|_R - \epsilon U_1 V_1, \qquad u_r\Big|_R = u_r^0\Big|_R + \epsilon U_1 P_1, \qquad \vu\Big|_{R'_o(\xi)} = 0
\end{equation}
where $V_1$ is the first associated Legendre Polynomial,  $\vu^0$ is the solution to the pure-treadmill problem with swimming speed $U_0$ provided in \citet{reigh2017two}, and $R'_o(\xi) = R_o\left[1-\epsilon \sum_{k} b_kP_k(\xi)\right]$
is the perturbed outer surface expanded based on Legendre polynomial and $b_k$ represent the amplitude of mode $k\geq 0$. {This is an alternative formulation of Eq. \ref{eq:surfacestroke} with $V_n(\nu)=2/\sqrt{3n(n+1)}K_n(\nu)$.} The new swimming speed is defined by $U = U_0 + \epsilon U_1$. We calculate how $U_1$ depends on the mode shapes of the outer surface perturbation. 

The first-order solution $\vu^1$ can be solved in the same manner as the original problem with a modified boundary condition at $r=R_o$ obtained from the Taylor-series approximation at the bounding surface $R'_o(\xi)$. The $\mathcal{O}(\epsilon)$ problem has updated boundary conditions:
\begin{equation}
    u_\theta^1\Big|_R = - U_1V_1,\qquad u_r^1\Big|_R = U_1P_1,\qquad u_\theta^1\Big|_{R_o} = \parD{u_\theta^0}{r}\Big|_{R_o} P_k,\qquad u_r^1\Big|_{R_o} = \parD{u_r^0}{r}\Big|_{R_o} P_k.
\end{equation}
Because we know $\vu^0$ analytically \citep{reigh2017two}, we can compute these derivatives exactly. Furthermore, we can use the expansion $u_\theta^1\Big|_{R_o} = \sum_{n=1}^\infty \gamma_n V_n$, and solving for the resulting coefficients using orthogonality which results in 
\begin{equation}
U_1 = \gamma_1\left[1- \frac{5}{3} \frac{\sum_{i=0}^{2} \lambda^i}{\sum_{i=0}^{4} \lambda^i}\right]
\end{equation}
where $\lambda=R_o/R$. The term in brackets is positive for all $\lambda$, so the adjustment to swimming speed depends entirely on the coefficient $\gamma_1$. We can use the orthogonality of associated Legendre polynomials to obtain the expression for $\gamma_1$ as:
\begin{equation}
    \gamma_1 =
    \, \frac{B_1 15 \pi^2\lambda}{8\,R(\lambda^5-1)} \left[-\left(\frac{\pi+4}{\pi}\right)^2 b_0 -\frac{5}{64} b_2 - \frac{9}{4096} b_4 + ...\right]
\end{equation}

The presence of $b_0$ mode corresponds with a simple decrease in the confining sphere's radius. This portion of the asymptotic result is consistent with a result from \citet{reigh2017two} that showed that increased confinement has a purely negative effect on squirmer swimming speed. The increased swimming speed $U_1$ will be the same sign as $\gamma_1$. This result shows that if the sphere is perturbed with the $0^\text{th}$ or even higher-order Legendre mode, the swimming speed decreases. Interestingly, this asymptotic result shows that the swimming speed can also result from the presence of the even Legendre modes in the perturbed outer surface. Depending on the sign of the modes $b_i$, these results show that perturbed surfaces can even lead to positive $U_1$.  

{A Legendre decomposition of perturbations to the yield surface of Fig.~\ref{fig:treadmill_flow} shows that the surface perturbations mostly consist of the second Legendre mode regardless of $\Bin$, For example at $\Bin{}=1$ , $\approx87\%$ of the surface Legendre mode energy is in the (negative) second mode. Therefore, the surface perturbation contributes to the \emph{increase} of the swimming performance, compared to the prediction based on the concentric spheres. This suggests that modifications to swimming speed come through two competing effects: the size and shape of the yield surface.}

\subsection{Multimode Squirmer}
As discussed above, in a Newtonian fluid, all modes after the first fail to produce propulsion. This is because Newtonian fluids are homogeneous and the boundary modes $\beta_n$, $n\geq 2$, are symmetric in a sense that ``forward'' components of the surface stroke motion exactly cancel out the ``backward'' components. It is reasonable to wonder whether the introduction of viscoplasticity could break this symmetry and, consequently, higher-order modes could contribute to locomotion. To see if this is the case in a Bingham fluid, we examine cases whose surface stroke motion comprises the first three modes. We visualize the results using a unit circle where each point corresponds to the $\beta_2$ and $\beta_3$ modes -- with all higher modes set to zero; this automatically sets the $\beta_1$ mode via the mode normalization. 

\begin{figure}
    \centering
    \begin{subfigure}[t]{0.49\textwidth}
        \centering
        \includegraphics[scale=0.85]{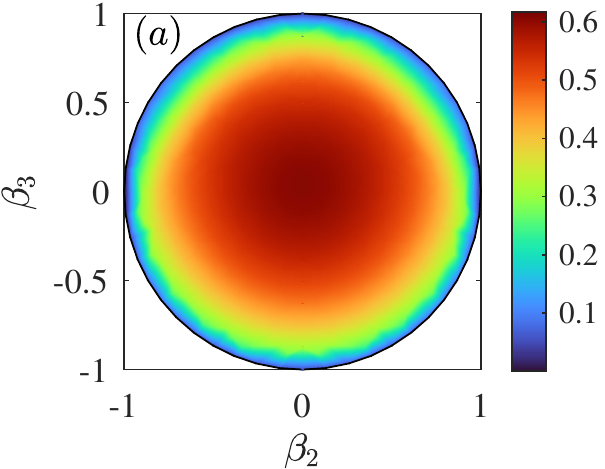}
    \end{subfigure}%
    \begin{subfigure}[t]{0.49\textwidth}
         \centering
        \includegraphics[scale=0.85]{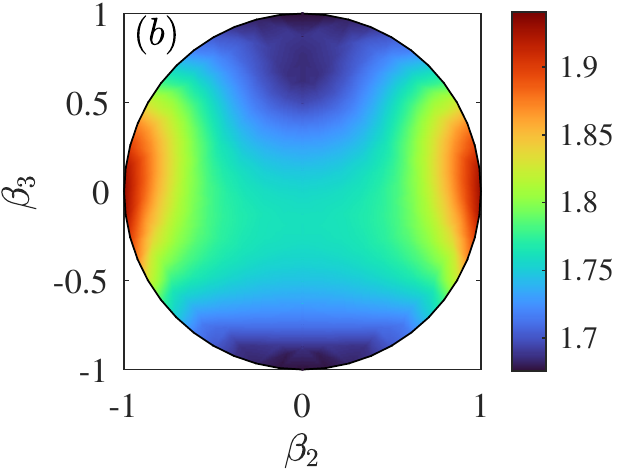}
    \end{subfigure}  
    
     \begin{subfigure}[t]{0.49\textwidth}
        \centering
        \includegraphics[scale=0.95]{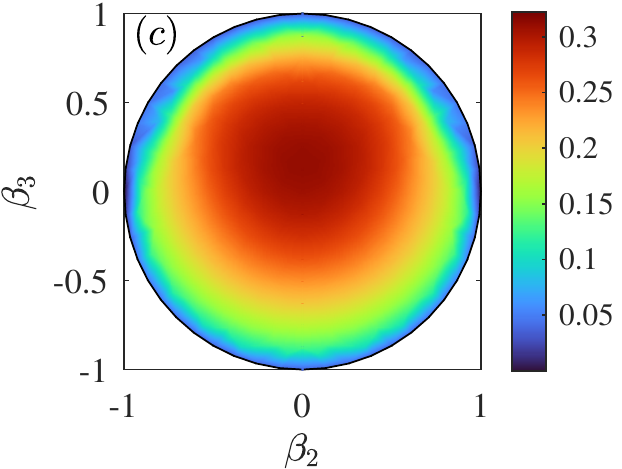}
    \end{subfigure}%
    \begin{subfigure}[t]{0.49\textwidth}
        \centering
        \includegraphics[scale=0.95]{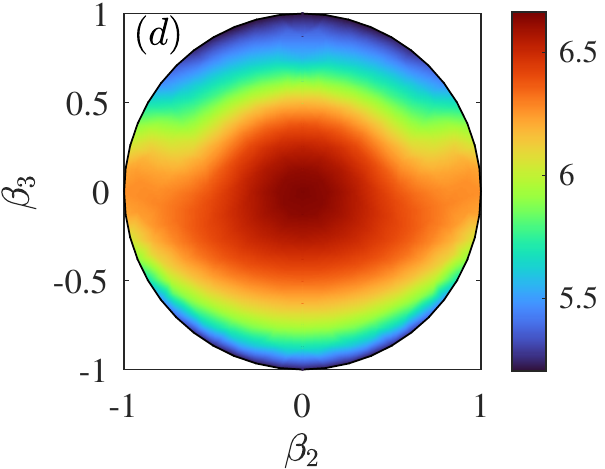}
    \end{subfigure}  
    
    \begin{subfigure}[t]{1\textwidth}
        \centering    \includegraphics[width=1\textwidth]{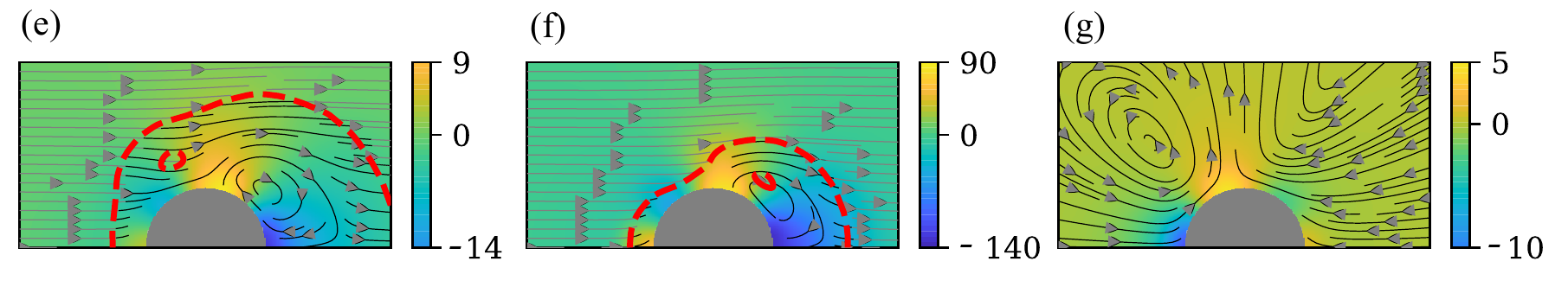}
    \end{subfigure}  
    \caption{{The effects of combinations of the first three modes a Bingham fluid on (a,c) swimming speed, $U$, and (b,d) normalized power expenditure, $\mathcal{P}^*$, for $\Bin{}=1$ (a,b) and $\Bin{}=10$(c,d). Flow fields at $(\beta_1, \beta_2, \beta_3) = (0, 0.79, 0.61)$,  are examined in (e) $\Bin{}=1$, (f) $\Bin{}=10$ and (g) Newtonian fluid. Streamlines represent the velocity field and color represents pressure. The yield surface is shown with red dashed line. The Bingham case can swim while the Newtonian case is immobile.}}
    \label{fig:multimode_global}
\end{figure}

\begin{figure}
    \centering

    \begin{subfigure}[t]{0.32\textwidth}
        \centering
        \includegraphics[scale=0.7]{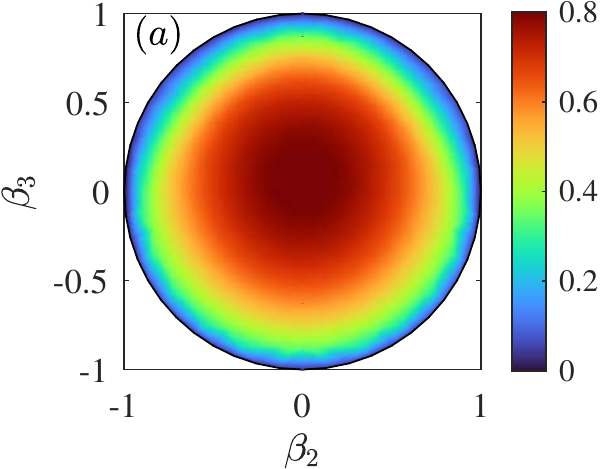}
    \end{subfigure}  
     \begin{subfigure}[t]{0.32\textwidth}
        \centering
        \includegraphics[scale=0.7]{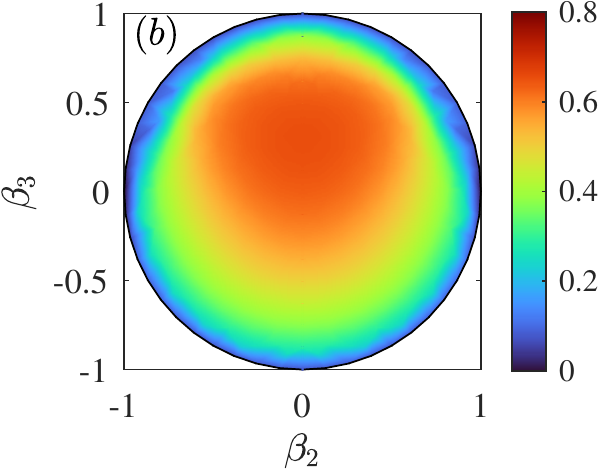}
    \end{subfigure}%
    \begin{subfigure}[t]{0.32\textwidth}
        \centering
        \includegraphics[scale=0.7]{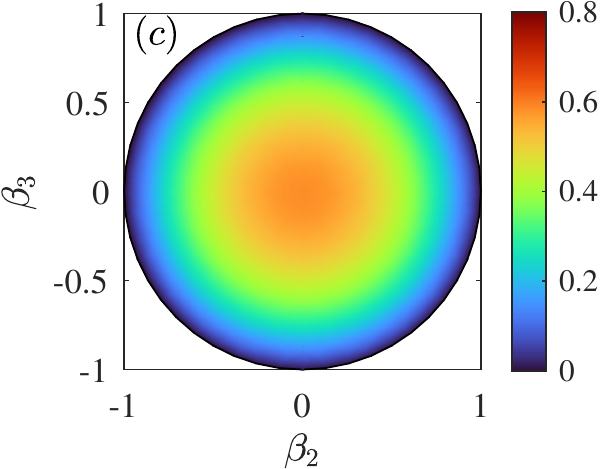}
    \end{subfigure}%
    \caption{{The swimming efficiency for the combinations of the first three modes in a Bingham fluid with $\Bin{}=1$ (a) and 10 (b) as well as a Newtonian fluid (c).}}
    \label{fig:multimode_globalEFF}
\end{figure}

\begin{figure}
        \centering
        \includegraphics[scale=0.8]{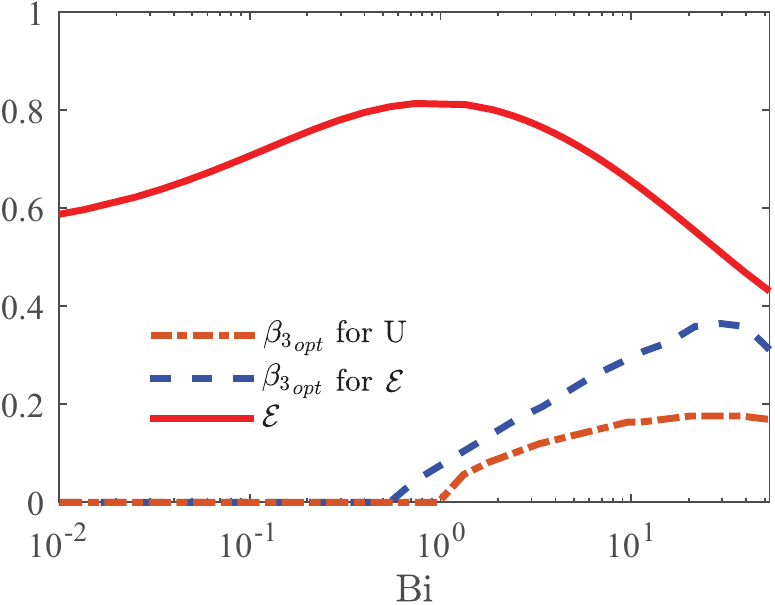}
    \caption{{The changes of maximum attainable swimming efficiency and optimal third mode actuation amplitude with $\Bin{}$. The solid line shows the maximum efficiency, the dashed line is the value of $\beta_3$ associated with maximum efficiency and the dash-dotted line is the value of $\beta_3$ where the maximum swimming velocity is observed.}}
    \label{fig:optimalmultimode_global}
\end{figure}

The results of swimming velocity and normalized power expenditure for $\Bin{}=1$ and $10$ are shown in Figure \ref{fig:multimode_global}. The results are largely consistent with those from the treadmill case in that the squirmer uses more energy in viscoplastic fluids and swims more slowly. However, an exception to this trend is when there are pure ``mixing modes''. In particular, when $\beta_3\neq 0$, there is non-zero swimming speed, a situation impossible in a Newtonian fluid. The fact that $\beta_2=1$ still results in no locomotion suggests that not all higher-order modes can lead to locomotion -- this is further explored in the next section. 
{Figure \ref{fig:multimode_global}b,d shows that the power expenditure behavior changes substantially depending on $\Bin{}$ value. In comparison, the power expenditure is mostly affected by the second mode $\beta_2$ for $\Bin{}=1$, while at a higher yield limit of $\Bin{}=10$, it mostly modifies with $\beta_3$  mode. Furthermore, interestingly, the highest swimming velocity is observed at a finite $\beta_3$ values rather than in pure treadmill swimming mode. In Fig. \ref{fig:optimalmultimode_global}, the optimal $\beta_3$ value associated with the maximum swimming velocity is shown versus $\Bin{}$. For low $\Bin{}<1$, the pure treadmill motion results in the highest swimming velocity, while for larger $\Bin{}$, the third mode of surface actuation enhances the locomotion of the swimmer. For very high yield limits of $\Bin{}>10$, the maximum swimming velocity always occurs at $\beta_3=0.17$. 

In Newtonian fluids, the addition of higher modes results in lower swimming efficiency as they do not contribute to the swimming speed and only add to the power expenditure by the swimmer \citep{stone1996propulsion,blake1971spherical}. In non-Newtonian fluids, however, previous studies observed that higher modes could be effective in swimming depending on the rheology of the fluid medium \citep{datt2015squirming,nganguia2017swimming}. Here, we examine how the yield limit of viscoplastic fluid affects the swimmer's efficiency with the surface stroke made up first three modes. Figure \ref{fig:multimode_globalEFF} shows the swimming efficiency of a squirmer for $\Bin{}=1$ (a), 10 (b), and Newtonian fluid (c). It is found that the swimmer has higher efficiency in yield fluids when $\Bin{}<30$ and the most efficient cases are associated with the combination of $\beta_1$ and $\beta_3$ modes. The region with high efficiency is shifted to higher $\beta_3$ modes when $\Bin{}$ increases. To explore the level of efficiency enhancement, in Figure \ref{fig:optimalmultimode_global}, 
the maximum attainable efficiency is shown as a function of the yield limit of Bingham fluid. Also, the corresponding $\beta_3$ magnitude that results in maximum efficiency is shown as a blue dashed line. It is also found (not shown here) that the inclusion of higher modes does not alter the results. The maximum efficiency is associated with pure treadmill motion for $\Bin{}<0.6$ while for a wide range of $0.6<\Bin{}<30$, the optimal third mode increases logarithmically with $\Bin{}$ according to $\beta_3 \approx 0.6+0.21 \text{log}_{10}\,(\Bin{}-0.6)$. The value of optimal $\beta_3$  reduces slightly for much larger $\Bin{}$ cases. The effect of the third mode on efficiency and its relation to the yield strength is reminiscent to previous observations in shear-thinning fluids where the efficiency change is interconnected to the relative magnitude of $\beta_2$ and $\beta_3$ modes \citep{nganguia2017swimming}.   }

In Figures \ref{fig:multimode_global}e-g, we examine the flow fields of the purely mixing mode $(\beta_1, \beta_2, \beta_3) = (0, 0.79,0.61)$ in two  Bingham environments with $\Bin{}=1$ and 10 as well as Newtonian fluid. The flow profiles of Bingham are quite different from Newtonian fluids; the Bingham fluid displays a stagnation point away from the surface in the aft body, which does not exist in the equivalent Newtonian fluid. This stagnation point is caused by the confinement effect of the yield surface. The next section examines pure higher-order modes to determine why the third mode can lead to locomotion.

\subsection{Higher pure modes}

In the Newtonian fluid, modes with $\beta_1=0$ produce zero swimming speed. As we saw in the previous section, this is not the case in a Bingham fluid so long as $\beta_3\neq 0$. To determine the extent of this phenomenon, we conduct simulations for $\Bin{}=0.1,1,10$ for all pure higher-order modes $\beta_n$, $n=1,\dots,12$. It is noticed that odd actuation modes can lead to swimming while even modes do not. Furthermore, this effect decreases with the mode number, e.g.~$\beta_3$ has a significantly higher swimming speed than $\beta_5$, which has a larger swimming speed than $\beta_7$, and so on. For all even modes, the swimming speed is always zero.

\begin{figure}
    \centering
     \begin{subfigure}[t]{1\textwidth}
        \centering
        \includegraphics[scale=0.75]{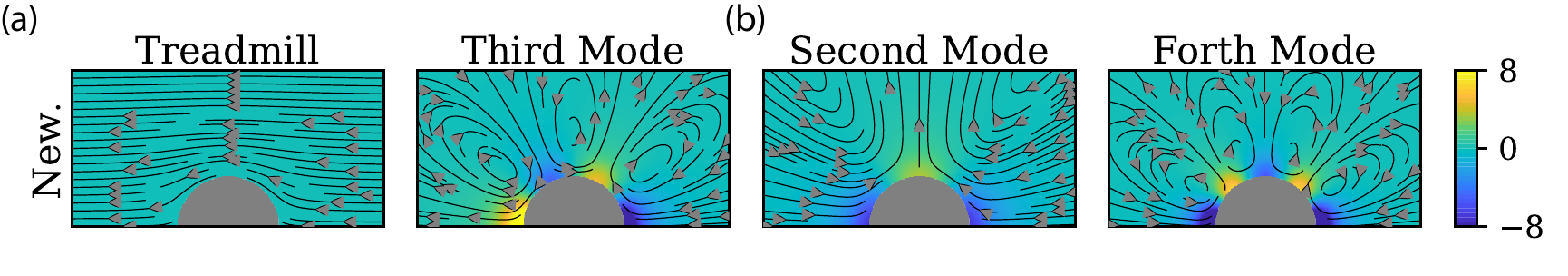}
    \end{subfigure}%
    
    \begin{subfigure}[t]{1\textwidth}
        \centering
        \includegraphics[scale=0.75]{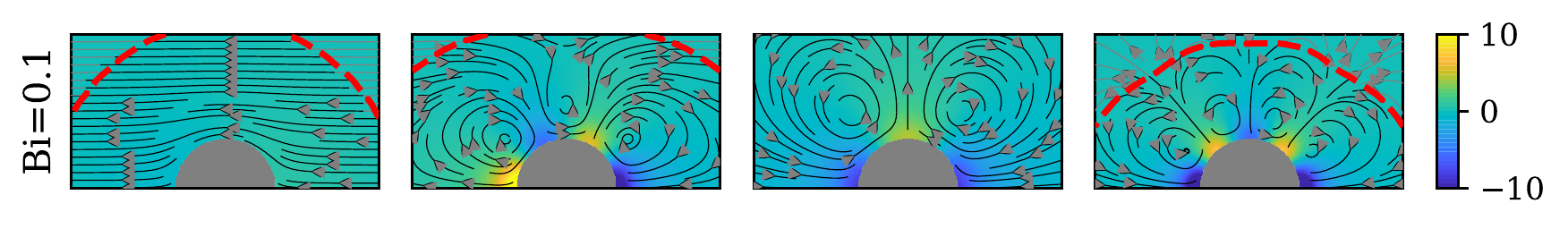}
    \end{subfigure}%
    
    \begin{subfigure}[t]{1\textwidth}
        \centering
        \includegraphics[scale=0.75]{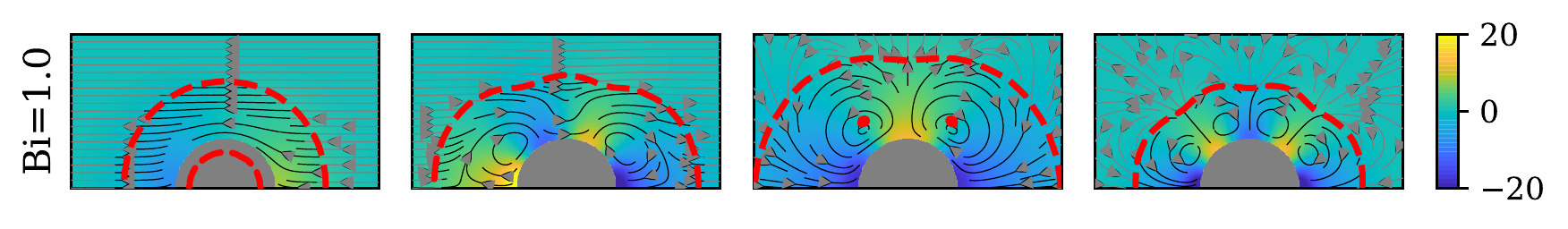}
    \end{subfigure}%
    
    \begin{subfigure}[t]{1\textwidth}
        \centering
        \includegraphics[scale=0.75]{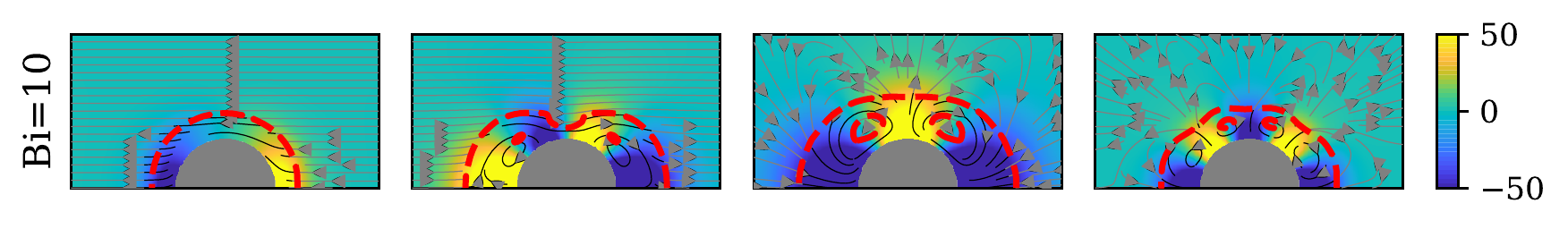}
    \end{subfigure}  
        
    \caption{Flow fields for odd (a) and even (b) modes. The odd modes display qualitative differences and maintain front-back anti-symmetry while the even modes do not display qualitative differences and maintain front-back symmetry. Streamlines represent kinematic velocity field, and color represents pressure field. the yield surface is shown with red dashed line. All squirmers shown in (a) are mobile and those shown in (b) are stationary. {The velocity magnitude along the gray lines outside the yield surface in (b) is several order of magnitude smaller than in the yield region, always smaller than $0.002$.} } 
    \label{fig:evenmodeflow}
\end{figure}

We examine the flow fields for odd (Fig.~\ref{fig:evenmodeflow}a) and even (Fig.~\ref{fig:evenmodeflow}b) modes in both a Bingham and Newtonian environment. The flow profiles for even modes are qualitatively similar between Bingham and Newtonian environments, while odd modes ($n\geq 3$) display qualitatively different flow profiles. Specifically, the only difference for even modes is that the vortices generated near the surface due to the boundary conditions shrink more tightly towards the body as the yield surface gets closer to the swimmer; this manifests as a confinement effect. On the other hand, odd-numbered modes, in particular mode $\beta_3$, have substantially different flow fields wherein a vortex presented in the Newtonian fluid disappears and a non-periodic flow path takes its place in the viscoplastic environment. We hypothesize that the confinement effect of the yield surface breaks up this vortex and, importantly, allows the top slip boundary condition to ``reach out'' to the far-field, thus generating non-zero propulsion. {This observation is aligned with prior results in shear-thinning fluids that the nonlinear rheology of the fluid medium can modify the near body flow and pressure field and results in new forms of non-Newtonian local and non-local forces on the body. More information about the contribution of the third stroke mode of the flow field around a squirmer in shear-thinning can be found in \cite{pietrzyk2019flow}.}

\begin{figure}
    \centering
    \begin{subfigure}[t]{0.45\textwidth}
        \centering
        \includegraphics[scale=0.75]{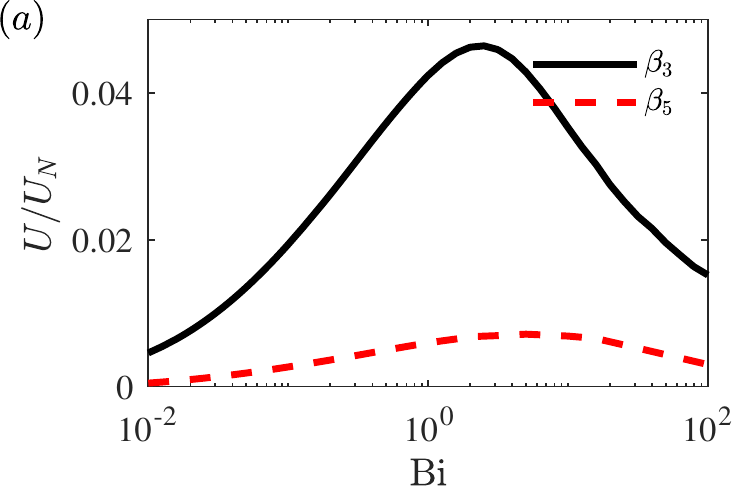}
    \end{subfigure}%
        \begin{subfigure}[t]{0.45\textwidth}
        \centering
        \includegraphics[scale=0.75]{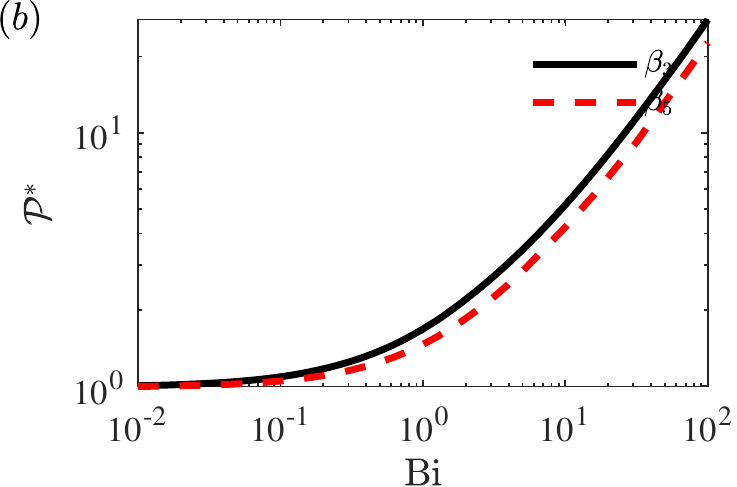}
    \end{subfigure}  
    \caption{Modes $\beta_3$ and $\beta_5$ generate non-zero swimming motion. An increasing power trend is also observed, similar to the treadmill case. Each odd-numbered mixing mode can generate its maximum swimming speed at a particular $\Bin{}$ range.}
    \label{fig:B3B5}
\end{figure}

Examining the effect of $\Bin{}$ on swimming speed at higher-order odd modes, we visualize the swimming speed (normalized by the swimming speed of a treadmill squirmer in Newtonian fluid) and power expenditure of pure $\beta_3$ and $\beta_5$ modes in Figure \ref{fig:B3B5}. It can be seen that, as $\Bin{}$ increases, both modes can achieve higher swimming speeds, although $\beta_5$ is growing significantly slower than $\beta_3$, the trend observed for higher-order odd modes. {Higher modes attain their optimal swimming speed at particular $\Bin{}$ values which increase with the mode number. For example, the maximum swimming speed of $\beta_3$ observed at $\Bin{}=2.2$ and it shifts to $\Bin{}=7$ for $\beta_5$.  Additionally, the power expenditure increases as $\Bin{}$ increases. This is related to the increasing confinement effect, which means the same kinematic slip velocity boundary conditions ``push harder'' against the surrounding fluid due to the close presence of plastic confinement.}

\section{Discussion}
\label{sec:discussion}

We have explored an axisymmetric squirmer in a Bingham fluid using the numerical simulation. It is shown that swimming in a Bingham fluid requires more power with the same kinematic boundary conditions for all modes. We argue that this effect is due to the increasing confinement effect from the yield surface. For the treadmill squirmer, we found that swimming speed decreased as the Bingham number $\Bin{}$ increased. Nevertheless, the shape of the yield surface can alleviate some of the decreasing effects. The efficiency of the swimmer, on the other hand, is higher at moderate $\Bin{}$ range before worsening at much higher Bingham numbers. {We found that specific pure higher-order modes, specifically odd-numbered modes, could achieve locomotion on their own, a result impossible in a Newtonian fluid. Also, the $\beta_3$ and $\beta_5$ modes' swimming speeds increase with increasing $\Bin{}$ for moderate Bingham numbers, an opposite trend compared to the treadmill mode. The swimming velocity and efficiency of the swimmer can be improved by including higher odd modes into the pure treadmill stroke (in particular, the third mode).   The combination of these results suggests that there is a ``crossover'' value of $\Bin{}$ at which the $\beta_3$ mode can be employed to enhance the efficiency and swimming speed compared to the treadmill mode.} This suggests the existence of different optimal strategies in stroke selection in viscoplastic fluids, similar to other types of non-Newtonian fluids, specially shear-thinning fluids \citep{pietrzyk2019flow,nganguia2017swimming}. 

The results here and previous studies on non-Newtonian modeled swimmers raise the following question that we invite the community to address. Many authors only consider the first two modes, $\beta_1$ and $\beta_2$, as it is believed that these sufficiently approximate the far-field flow dynamics of real microswimmers. Our results suggest that investigating higher-order modes is important for discovering possible alternative optimal swimming strategies in complex fluids, specially viscoplastic and elastoviscoplastic fluids. We believe it is worth examining existing analytic results in viscoplastic fluid involving self-induced confinement with a focus on higher-order modes; e.g.~see \citet{reigh2017two} and  \citet{nganguia2020squirming}. {The nonlinear interaction between the modes is also an interesting future extension of this study.}

Finally, we address the direct applicability of these results to \emph{H.~Pylori}. The Bingham fluid model ignores an important property of mucus, namely the fact that it is thixotropic. Therefore, our steady analysis presented here should only be considered an approximation, and we have possibly missed temporal effects present in real stomach mucus. Additionally, typical scales for \emph{H.~Pylori} include a length of $R=\SI{1}{\mu\meter}$, swimming speed of $V=\SI{10}{\mu\meter\per\second}$, and characteristic viscosity of gastric mucus of $\mu_0=\SI{0.05}{\pascal\second}$ \citep{curt1969viscosity}.
{Unfortunately, we could not find an estimate for the yield stress of gastric mucus, but using a range of estimates for common viscoplastic materials from ketchup ($\SI{15}{\pascal}$) to hair gel ($\SI{135}{\pascal}$) provides an estimated range for $\Bin{}$ of 30 -- 270. As our results suggest that higher-order odd modes become more effective at driving microswimmer propulsion as higher $\Bin{}$, this insinuates the possibility that \emph{H.~Pylori} incorporates a modified mode of locomotion to more effectively swim through the stomach mucus.} Related to this, results of two-dimensional squirmers showed that coupling treadmill and mixing modes achieved maximum propulsion at an intermediate $\Bin{}$ value \citep{supekar2020translating}. Finally, real gastric mucus is chemically changed by the bacteria itself. Therefore, gastric mucus should not be treated as a homogeneous viscoplastic material regarding \emph{H.~Pylori} locomotion -- a more realistic model could include spatially-varying yield stress $\yieldstress$ subject to the spatial distribution of an emanating chemical.

\bibliography{main}

\end{document}